\documentclass[aps,twocolumn,prb]{revtex4}
\usepackage{epsfig,amsmath,graphics}
\usepackage{bm}
\begin{document}

\title{Lagrange multiplier based transport theory 
for quantum wires}

\author{D.S. Kosov}


\affiliation{Institute of Physical and Theoretical Chemistry, \\ 
J. W. Goethe University, Marie-Curie-Str.~11, D-60439 Frankfurt, Germany}
\begin{abstract}
We discuss how the Lagrange multiplier method of non-equilibrium 
steady state statistical mechanics can be applied to describe
the electronic transport in a quantum wire. We describe 
the theoretical scheme using tight-binding model.
The Hamiltonian of the wire is 
extended via a Lagrange multiplier to ``open'' the quantum 
system and to drive  current through it. The diagonalization 
of the extended Hamiltonian yields  the transport 
properties of  wire. We show that
the Lagrange multiplier method is  equivalent to the 
Landauer approach within the considered model.
\end{abstract}
\maketitle

\newpage
In recent years, several experimental groups have reported
the measurements of the transport properties of carbon nanotubes, 
self-assembled monolayers of conjugated polymers and even  
individual molecules \cite{joachim2000}.  This development has attracted much 
attention from theoreticians and most theoretical works reported 
so far are based on  the Landauer type theory \cite{landauer70}.
Within the Landauer approach  it is assumed that
incoming electrons are scattered along the molecular wire and the conductance 
can be directly evaluated from the scattering T-matrix. The nonequilibrium 
Green's functions or 
Lipmann-Schwinger equation approaches
have been used to compute $T$-matrix \cite{xue2003,galperin2002,diventra2001}. 
This extensive calculation work on molecular 
junctions have provided basic understanding of the physical processes 
although the aim of the 
direct reproducing of  experimental I-V characteristics 
is still elusive \cite{nitzan2003}. 
This current status of 
electron transport calculations in molecular wires
requires that other, ``non-Landauer-type'' theoretical descriptions 
should  also be properly explored. 

A quantum wire with direct, static current is
a nonequilibrium  steady state system. This can be understood based upon
the following qualitative discussion. Suppose that the left end of the wire 
works as a equilibrium source of the electrons and the right end works 
as a drain. The source and drain have different time-independent
chemical potentials  ($\mu_s > \mu_d$).
The source pumps electrons
into the wire  trying to establish the equilibrium, but at the
same time the drain keeps pulling electrons
out of it also trying to equilibrate the wire with itself.  
The equilibrium is never achieved as long as the chemical potentials
$\mu_s $ and $ \mu_d$ are kept fixed or if the current is constrained
to be time-independent constant. The system stays in nonequilibrium 
steady state provided that there is  the time-independent current through it.

Having established that a quantum wire with direct current is a 
steady state nonequilibrium quantum system, we  can apply the 
powerful machinery of  modern theoretical methods available 
in nonequilibrium steady state statistical mechanics to describe 
it. Recently, there have been  considerable advances towards 
the formal 
understanding of nonequilibrium steady 
state open quantum systems on the basis of the 
Lagrange multiplier based method 
\cite{antal97,antal99,kosov2002,eisler2003}.
The basic assumption of the  Lagrange multiplier based method is that
the origin of  current is irrelevant to  physics of the steady state
or in other words that
a homogeneous  current-carrying state is the same whether it 
is introduced by the reservoirs, i.e. via boundary conditions,
or by a bulk driving field. 
A nonequilibrium steady state can be produced by the following three-step
``algorithm''\cite{antal97,kosov2002}:
\begin{enumerate}
\item
Define the operator of  current  $J$ via the continuity equation.
\item
Extend the Hamiltonian by adding the term ($ -\lambda J$) where $\lambda$
is a Lagrange multiplier.
\item 
Diagonalize the extended Hamiltonian.
\end{enumerate}

The aim of this paper is to place 
electronic transport calculations into the general framework of
the modern development of  nonequilibrium steady state statistical 
mechanics. We begin with the model tight-binding Hamiltonian 
for an infinitely long quantum wire
\begin{equation}
H_0= -t \sum_{n \sigma} (a^{\dagger}_{n+1 \sigma} a_{n \sigma} 
+ a^{\dagger}_{n \sigma} a_{n+1 \sigma})
\; ,
\end{equation}
where $t$ is the hopping matrix element. The operators
$a^{\dagger}_{n \sigma}$  ($a_{n \sigma}$) create
(annihilate) single electron on the site $n$,
where $\sigma= \pm \frac{1}{2}$ is the spin of the electrons. 
Being fermionic operators the operators $a^{\dagger}_{n \sigma}$  
($a_{n \sigma}$) obey
the standard anticommutation relations:
  \begin{eqnarray}
 &&\{ a_{n \sigma}, a^{\dagger}_{m \sigma'} \} 
 = \delta_{nm} \delta_{\sigma \sigma'}
 \;, \\
 && \{ a^{\dagger}_{n \sigma}, a^{\dagger}_{m \sigma'} \} = 
 \{ a_{n \sigma}, a_{m \sigma'} \} =0 \;.
 \end{eqnarray}
The left part of the wire is considered to be source for current 
and the right end serves as  drain. We do not use any assumptions
regarding the physical nature of the source/drain  we merely assume 
they exist and are able to
maintain a steady, i.e. time-independent, current through the wire.

As the first step, we define the operator of  current via the 
continuity equation. 
The number of electrons on the site $n$ is given by the 
expectation value of the
operator 
\begin{equation}
N_n = \sum_{\sigma} a^{\dagger}_{n \sigma}  a_{n \sigma}\;.
\end{equation}
By making use of Heisenberg representation
 the continuity equation can be written as 
the Heisenberg equation-of-motion for the operator $N_n$:
\begin{equation}
\dot{N_n} = i \left[ H_0, N_n \right] \; .
\label{cont-eq-1}
\end{equation}
Given the standard anticommutation relations between electron 
creation and annihilation operators, the r.h.s. commutator 
(\ref{cont-eq-1}) can be readily computed:
\begin{eqnarray}
&&\dot{N_n} = - i t \sum_{\sigma} (a^{\dagger}_{n+1 \sigma} a_{n \sigma} 
- a^{\dagger}_{n\sigma} a_{n+1 \sigma})
\nonumber
\\
&&+ i t \sum_{\sigma} (a^{\dagger}_{n \sigma} a_{n-1 \sigma} 
- a^{\dagger}_{n-1 \sigma} a_{n \sigma})\;. 
\label{cont-eq-2}
\end{eqnarray}
Comparison of  eq.(\ref{cont-eq-2}) with the finite difference 
expression for continuity equation
$\dot{N_n} = -(j_n -j_{n-1})$  
yields the definition of the operator of current through the site $n$:
\begin{equation}
j_n = i t \sum_{\sigma} (a^{\dagger}_{n+1 \sigma} a_{n \sigma} - 
a^{\dagger}_{n \sigma} a_{n+1 \sigma})\;.
\nonumber
\end{equation}
By making the sum of  on-site currents along the wire we obtain 
the net current through the wire
\begin{equation}
J = i t \sum_{n\sigma} (a^{\dagger}_{n+1 \sigma} a_{n \sigma} - 
a^{\dagger}_{n \sigma} a_{n+1 \sigma})\;.
\label{current}
\end{equation}

The next step in our scheme is to fix the net current via a
Lagrange multiplier $\lambda$.  To this end the Hamiltonian $H_0$  
is modified by adding the term which constraints the macroscopic 
current $J$:
\begin{eqnarray}
&& H=H_0 - \lambda J\;= 
- t \sum_{n \sigma} (a^{\dagger}_{n+1 \sigma} a_{n \sigma} + 
 a^{\dagger}_{n \sigma} a_{n+1 \sigma})
\nonumber
\\
&&-\lambda i t \sum_{n\sigma} (a^{\dagger}_{n+1 \sigma} a_{n \sigma} -
 a^{\dagger}_{n\sigma} a_{n+1 \sigma})\;.
\label{h0-lambdaj}
\end{eqnarray}
The Hamiltonian (\ref{h0-lambdaj}) is hermitian although it is no longer 
a real operator. The term $(-\lambda J)$ breaks the symmetry 
between electrons moving along the wire in opposite directions. Now
all physical properties of the system including transport characteristics
should be obtained from the diagonalization of the Hamiltonian $H $ not $H_0$.

The final step is the
diagonalization of the extended Hamiltonian. 
Using the periodic boundary conditions $ a_{n\sigma} = a_{n+N \sigma}$ 
($N$ -the ''length'' of the box)
the Hamiltonian (\ref{h0-lambdaj}) 
can be diagonalized via the Fourier transformation
\begin{equation}
  a_{n\sigma} =\frac{1}{\sqrt{N}} \sum_k \exp(-ikn) a_{k \sigma}\; ,
\label{transform}
\end{equation}
where the sum  over the k runs over the first Brillouin zone
($k=\frac{2\pi}{N} i$ with $i=0,....,N-1$). 
Furthermore, the operators 
$a^{\dagger}_{k\sigma}$($a_{k\sigma}$) 
still obey the fermionic anticommutation relations.
The transformation (\ref{transform})
brings the Hamiltonian (\ref{h0-lambdaj})
to the diagonal form 
\begin{equation}
H=\sum_{k\sigma} E(k) a^{\dagger}_{k \sigma} a_{k \sigma} \;,
\end{equation}
with the current-dependent dispersion relation 
for the quasiparticle energy (band energy): 
\begin{equation}
E(k)=
-2 t (\cos(k) + \lambda \sin(k)) \;.
\label{dispersion}
\end{equation}
As we let the current tend to zero, i.e. $\lambda \rightarrow 0 $, 
we recover the usual result for the band energy 
$ E(k)=-2 t \cos(k)$.
The  dispersion relation
(\ref{dispersion}) is not yet in a form 
allowing for the energy to be computed as the Lagrange multiplier  
$\lambda$ is not known yet.
The additional equation for the Lagrangian multiplier
$\lambda$ is obtained if  
the density of the 
expectation value of the net current operator (\ref{current}) 
over the ground state of the
Hamiltonian (\ref{h0-lambdaj}) 
\begin{equation}
\frac{1}{N} \langle J \rangle = \frac{4 t}{N} \sum_{k} \sin(k) n_k
\label{<J>}
\end{equation}
is required to yield the {\it a priori} known current density $I$. 
The occupation numbers $n_k$  equals to 1 if the band energy $E(k) \le 0 $ 
and zero otherwise for half-occupied conductance band, 
i.e. $n_k =\theta(-E(k))$
where $\theta$ is the Heaviside step function.

We finally demonstrate application of the method with numerical 
and analytical examples.
The numerical calculations were carried out 
for half-occupied conductance band  with the box length $N=200$.  
The value of the hopping integral $t$ is chosen to be 2.5 eV which is the 
standard value for monoatomic metallic wires \cite{heeger88}. 
The Lagrange multiplier $\lambda$ is obtained via the numerical solution 
of the following nonlinear equation:
\begin{equation}
  \frac{4t}{N}  \sum_{k} \sin(k) n_k = I, 
\label{system}
\end{equation}
\begin{figure}
  \centerline{
\epsfig{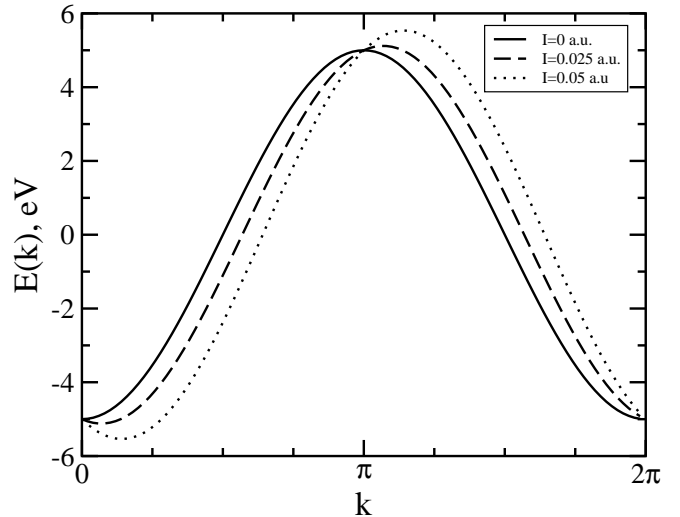}}
\caption{
The current-dependent band energies of the one dimensional
conductor for various current densities:
solid line - zero current, dashed - 0.025 a.u. and dotted
-0.05 a.u.
}
\label{fig:band}
\end{figure}

In Fig.1, the current-dependent 
band energies are shown for three different values of the current density. 
Since the commutator $[H,J]$ vanishes exactly it is perfectly
possible to characterize each electronic $k$-state by the 
value of the current in this state:
\begin{equation}
J_k = \langle a_{k\sigma} J  a^{\dagger}_{k\sigma} \rangle =2t \sin(k).
\end{equation}
It enables us to distinguish the electronic states by their
negative or positive contributions to the net current.
From Fig.1 we see that 
the current re-arranges the band structure in such a way 
that the part of the
band which corresponds to  states with positive current ($0<k<\pi$) 
has lower energy 
than the zero-current band while the energy of  states with  
negative current
are increased with respect to the zero-current solution. This selective
to current alternation of the band energy splits the occupation 
numbers of  states with  positive and negative $J_k$.

Now we turn our attention towards how the voltage drop
can be defined within the Lagrange multiplier based 
transport theory. For the most regimes observed in molecular wires  the
electrons are transmitted through the levels in the vicinity of the 
Fermi-energy $E_f$. Therefore to associate 
voltage with current we need to get clear physical picture
of Fermi-energy response upon current. 
In the first Brillouin zone there are always two 
$k$ which correspond to the Fermi-energy
\begin{equation}
E_f = -2t(\cos(k) +\lambda \sin(k)).
\label{fermi}
\end{equation}
 First solution of eq.(\ref{fermi}), $k_+$  ($0 < k_+ \le \pi$),
gives a positive contribution to total net current while  
the second solution, $k_-$ ($\pi < k_- \le 2\pi$), contributes negatively.
By turning off the current (i.e. in the limit of $\lambda=0$) the energy of 
of $k_+$-state is increased
by the value of $(-2t\lambda \sin(k_+))$ while the energy 
of $k_-$-state  goes down
by $(-2t\lambda \sin(k_-))$.
In the terminology of the Landauer theory for reflectionless 
contact \cite{datta95}, 
the energies of states  originating from the source, i.e. $k_+$ states,
are effectively decreased by voltage while    
the energies of states 
occupied by electrons  from  the drain, 
i.e. $k_-$ states, are increased by voltage.
Given that the voltage does the same job as the term $(-2t\lambda \sin(k_-))$
we arrive to the following equations: 
\begin{eqnarray}
&&E_f= -2t\cos(k_{\pm}) -2t\lambda \sin(k_{\pm}))
\nonumber
\\
&&\equiv -2 t \cos(k_{\pm}) 
\mp \frac{1}{2} U,
\label{fermi2}
\end{eqnarray}
where $U$ denotes the voltage drop. Eq.(15) 
yields the following expression for the voltage $U$:
\begin{equation}
U=2t\lambda (\sin(k_+) -\sin(k_-))
\label{voltage}
\end{equation} 
The voltage produces splitting ( as in standard Landauer theory) 
the 
Fermi-Dirac occupation numbers of the the states 
with positive and negative $J_k$.
The Fermi-Dirac occupation numbers are given in Fig.2 for two
different values of the currents: $I=0.01$ a.u. (the lower panel)
and  $I=0.025$ a.u. (the upper panel). The Heaviside step
function is smeared out as the corresponding Fermi-Dirac function 
taken at temperature $T=0.3$ eV. The values of the voltage 
computed  by the formula (\ref{voltage}) are also plotted on Fig.2.
\begin{figure}
  \centerline{
\epsfig{figure=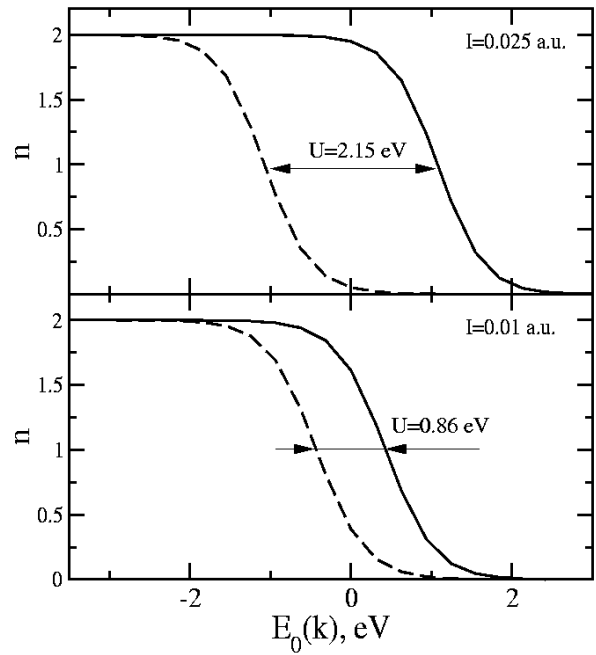,
width=1\columnwidth
}}
\caption{
 The Fermi-Dirac occupation numbers of
current carrying states. The solid line is for the states 
with positive contribution to the net current ($J_k>0$).
The dashed line is for  
the occupation probabilities of  the states with negative $J_k$. 
$E_0(k) = -2t\cos(k)$ is the band energy of the zero-current wire.
}
\label{fig:occ}
\end{figure}
%
%
We would like to emphasize that 
the definition of  voltage described above is 
applicable directly only if current  $J$ is an
integral of motion for the Hamiltonian $H_0$, i.e. $[H_0,J]=0$.
The commutator $[H_0,J]$ does not necessarily
vanish exactly in general case and the occupation numbers of
``upstream'' and ``downstream'' electrons are tangled
because the eigenstates of the extended Hamiltonian $H$ are 
no longer  uniquely characterized by  value of the current 
in these states.
Therefore it might be practically cumbersome to extract the voltage 
for a real molecular wire by looking at the populations of states 
carrying currents in the different directions.
The rigorous definition of the applied voltage
is generally  a very controversial issue if one does not invoke 
to noninteracting electron reservoirs  to 
represent boundary conditions 
and similar problems are encountered in ref.\cite{johnson95}.
One can always resort to the exact definition of the voltage
based upon the following physical picture:
a voltmeter determines the voltage drop 
along the wire by measuring
the work required to move a point unit charge from one side of 
the wire to the other. This brute force algorithm of calculating 
the work done on a test point charge can be always applied although it is
not computationally the most tractable scheme. The evaluation of the 
different voltage definitions for the correlated electrons will be 
discussed elsewhere.

To demonstrate a compatibility with the Landauer approach
we re-derive the Landauer expression for the conductivity 
\cite{landauer70} using our formulae
for the net current (\ref{<J>}) and our definition of the 
voltage \ref{voltage}.
Sum (\ref{<J>}) can be replaced by the integral provided that
for large $N$ the 
spectrum $E(k)$ can be considered as a quasi-continuum:
\begin{eqnarray}
&& I =  \frac{2t}{\pi} 
\int\limits_0^{2\pi} dk  \sin(k) n_k 
\nonumber
\\=
&&\frac{2t}{\pi}
\int\limits_0^{\pi} dk  \sin(k) (n_k - n_{k+\pi}).
\label{integral1}
\end{eqnarray}

Assuming the small current, i.e. small $\lambda$,  and by making
use of the Taylor expansion the following expression
for the occupation numbers difference can be found
\begin{equation}
n_k - n_{k+\pi} = \theta(\cos(k))-\theta(\cos(k+\pi))
+ 2  \lambda \sin(k) \delta(\cos(k)).
\nonumber
\end{equation}
Only the $\delta$-function term gives a 
non-vanishing contribution to the integral (\ref{integral1}) 
and the direct integration results into the following simple
expression for the current density:
\begin{equation}
I =  \frac{4t \lambda}{\pi}\;, 
\label{I}
\end{equation}
To compute voltage by the formula (\ref{voltage}) $k_+$ and $k_-$
should be determined. For half-occupied band $E_f=0$ and the 
eq.(\ref{fermi}) can be straightforwardly resolved:
\begin{equation}
k_+=\pi-\arctan(\frac{1}{\lambda}) \;\;\;\;\;\;\; k_-=2\pi-
\arctan(\frac{1}{\lambda})
\end{equation}
Then eq.(\ref{voltage}) yields the voltage
\begin{equation}
U= \frac{4t\lambda}{\sqrt{\lambda^2+1}}=4t\lambda+O(\lambda^2)
\label{U}
\end{equation}
Assuming again that the current is small and
substituting the expressions for current (\ref{I}) 
and voltage (\ref{U}) into the 
definition of conductance $ G=I/U$ we obtain:
\begin{equation}
G= \frac{1}{ \pi  } \;,
\label{landauer2}
\end{equation}
which is the Landauer value of G for a single transport channel of 
an ideal one-dimensional lead \cite{datta95}.
Re-deriving the standard result
we demonstrate that our method is equivalent 
to the Landauer approach within the considered model. 
It should be noted, however, 
that there is no guarantee that this formalism will exactly
reproduce the Landauer results in more complicated cases.
 
%
%
Finally, we discuss how the Lagrange multiplier based transport theory
 can be applied to a molecular wire with inhomogeneous
electron density. To describe this realistic scenario we 
specify the constraint on 
the current density distribution in 
the following form (constrained continuity equation):
\begin{equation}
\int dy\, dz\, j_x({\bf r})= I,
\end{equation}
where $I$  is a desired value of the current through the wire.
Within this description, net current flow is aligned along
the $x-$axis  and the net current flow across a cross section
$\int dy\, dz\, j_x({\bf r})$ is constrained, and this
quantity is readily available experimentally. It leads to
the following extended Hamiltonian:
\begin{equation}
H =H_0 - \int dx \lambda(x) \int \int dy\, dz\, j_x({\bf r}),
\end{equation}
where $\lambda(x)$ is the pointwise Lagrange multiplier in 
the additional space- and current-dependent term. 
An inhomogeneous interacting electron gas has been considered  
based upon a variational analog of the Lagrange multiplier based transport
theory and within the density functional theory
one generally needs to solve the set of integral self-consistent equations for
the pointwise Lagrange multiplier $\lambda(x)$ and the 
current-carrying Kohn-Sham orbitals \cite{kosov2003}. 

The final comment regarding the general applicability of 
the method is in due order.
The common characteristic feature of all nonequilibrium 
steady state systems 
is  the constant flux of a certain physical quantity, 
e.g. heat conduction (energy current), 
diffusion (particle current), electrical conductivity (charge current). 
We have discussed the scheme on an example of tight-binding Hamiltonian but
it can be extended with a slight modification to
any open steady state quantum system, e.g. the Ising model with 
the energy flux \cite{eisler2003}. Constant current 
Hartree-Fock and Kohn-Sham approximations have also been recently 
formulated within the approach \cite{kosov2001,kosov2003}.

In this paper, we have given the Lagrange multiplier 
based formulation of  electronic transport problem. 
We discussed the three-step practical algorithm
to produce current carrying steady states in quantum wires. 
First, we defined the operator of  macroscopic current 
via continuity equation. Next, 
tight-binding Hamiltonian was modified by the Lagrange multiplier 
term to account for the steady current. Then, the 
current-dependent band-energies and the occupation numbers 
were obtained by the diagonalization of the extended 
Hamiltonian. The definition
of the applied voltage which is compatible with 
the Landauer description was discussed.
A sample calculation on an albeit simple model system produced exact
agreement with results obtained from the Landauer theory.  
While the one-dimensional tight-binding model avoids a number of 
computational difficulties in using this approach, it shows how a 
Lagrange multiplier
can be applied to describe electronic transport properties of quantum wires. 

\acknowledgements 

The author wishes to thank A.~Nitzan for stimulating discussions and for
critical reading the manuscript.



\begin{thebibliography}{50}

\bibitem{joachim2000} C.Joachim, J.K. Gimzewski and A.Aviram, 
Nature {\bf 408}, 541 (2001). 

\bibitem{landauer70} R.Landauer, Philos. Mag. {\bf 21}, 863 (1970).

\bibitem{xue2003}
Y.Xue and M.A. Ratner, Phys. Rev. B  {\bf 68} , 115406 (2003)

\bibitem{galperin2002} M. Galperin, S. Toledo and A. Nitzan,
J. Chem. Phys. {\bf 117}, 10817 (2002).

\bibitem{diventra2001} M. Di Ventra, S.T. Pantelides, N.D. Lang, 
Phys.Rev.Lett. {\bf 84}, 979 (2000).

\bibitem{nitzan2003} A. Nitzan and M.A. Ratner, 
Science {\bf 300}, 1384 (2003).

\bibitem{antal97} T.Antal, Z.R\'acz, and L.Sasv\'ari,
Phys.Rev.Lett. {\bf 78}, 167 (1997).

\bibitem{antal99} T.Antal, Z.R\'acz, A.R\'akos, and G.M.Sch\"utz,
Phys.Rev. E, {\bf 59}, 4912 (1999).

\bibitem{eisler2003} V.Eisler, Z.R\'acz, and F. van Wijland, 
Phys.Rev. E, {\bf 67}, 056129 (2003).

\bibitem{kosov2002} D.S.Kosov, J.Chem.Phys., {\bf 116}, 6368 (2002).


\bibitem{heeger88}A. J. Heeger, S. Kivelson, J. R. Schrieffer, and W.-P. Su, 
Rev.Mod.Phys. {\bf 60},  781 (1988). 

\bibitem{datta95} S.Datta, Electronic transport in mesoscopic systems,
Cambridge University Press, 1995.

\bibitem{johnson95} M.D.~Johnson and O.~Heinonen, Phys.Rev. B, {\bf 51},
14421 (1995).

\bibitem{kosov2001} D.S.Kosov and J.C.Greer, 
Phys. Lett. A {\bf 291}, 46 (2001).

\bibitem{kosov2003} D.S.Kosov, J.Chem.Phys., {\bf 119}, 1 (2003).



\end{thebibliography}
\end{document}